\documentclass[%
 reprint, superscriptaddress,
 amsmath,amssymb,
]{revtex4-2}

\usepackage[english]{babel}
\usepackage{graphicx}    
\usepackage[caption=false]{subfig}
\usepackage{siunitx}
\usepackage{hyperref}
\usepackage{bm}

\DeclareSIUnit\electronmass{\text{\ensuremath{m}}_{\mathrm{e}}}
\DeclareMathAlphabet\mathbfcal{OMS}{cmsy}{b}{n}

\DeclareSIUnit\angstrom{\text{\AA}} 

\begin{document}

\preprint{APS/123-QED}

\title{\texorpdfstring{Interband transition orbit probed in de Haas-van Alphen\\ oscillations in the (double) Dirac semimetal NbTe$_4$}{}}

\author{Maximilian Daschner}
\email{md867@cam.ac.uk}
\email{maximilian.daschner@lmu.de}
\affiliation{Cavendish Laboratory, University of Cambridge, Cambridge CB3 OHE, United Kingdom}
\affiliation{Fakultät für Physik, Ludwig-Maximilians-Universität, München, Germany}

\author{F. Malte Grosche}
\email{fmg12@cam.ac.uk}
\affiliation{Cavendish Laboratory, University of Cambridge, Cambridge CB3 OHE, United Kingdom}


\begin{abstract}
NbTe$_4$ undergoes multiple charge density wave transitions that have attracted great interest in this material for decades. Previous work has shown that the crystal obtains the space group \textit{P4/ncc} (130) at temperatures below \SI{50}{\kelvin} which allows for the existence of eightfold degenerate double Dirac points in the band structure. We provide insights into the electronic structure of this material through density functional theory (DFT) calculations, and a rotation study of de Haas - van Alphen (dHvA) oscillations in the magnetic torque. We find that NbTe$_4$ exhibits magnetic breakdown orbits between electron and hole pockets.

\end{abstract}

\maketitle

\section{Introduction}
Transition metal chalcogenides (TMC) have recently gained interest due to their exotic topological properties such as the presence of Weyl nodes in WTe$_2$ \cite{li2017evidence} and MoTe$_2$ \cite{deng2016experimental}, Dirac nodes in ZrTe$_5$ \cite{yuan2016observation}, Dirac nodal lines in ZrSiX (X=S,Se,Te) \cite{singha2017large,hu2016evidence}, and their potential in hosting axionic charge density waves in (TaSe$_4$)$_2$I \cite{gooth2019axionic}. Besides their topological band structures, TMCs such as the sister compounds TaTe$_4$ \cite{bjerkelund1964crystal,bjerkelund1964properties,boswell1987phase,bennett1992incommensurate,sambongi1993shubnikov,prodan1995charge,luo2017resistivity,gao2017anisotropic,sun2020discovery,zhang2020eightfold,yuan2020pressure,liu2021first,guster2022competition,rezende2023experimental,xu2023orbital,zhang2023charge,nataj2023phonon,rong2024dominant,rezende2024experimental,bjerkelund1968high,boswell1983charge,boswell1984charge,mahy1982evidence,walker1985model,eaglesham1985microstructural,rouxel2012crystal,tadaki1990electrical,bennett1991observation,bennett1992effects,kusz1995modulated,butz2013nuclear,boswell2012advances,nataj2024raman} and NbTe$_4$ \cite{selte1964crystal,selte1965magnetic,mahy1983electron,bullett1984pd,whangbo1984band,sahu1985possible,mahy1985direct,bohm1985modulated,boswell1986structural,mahy1986microstructure,van1986determination,morelli1986transverse,ikari1987electrical,bohm1987high,van1987dualistic,prodan1987dualistic,walker1988nbte,morelli1989analysis,morelli1989novel,chen1989superspace,prodan1990structures,goff1991order,kusz1993incommensurate,coluzza1993high,kusz1994low,van1995incommensurate,prodan1998scanning,duvsek2001refinement,mori2003low,stroz2004applied,wu2009controllable,yang2018pressure,de2019large,galvis2023nanoscale,wu2023superconductivity,petkov2023charge,shuang2023nbte4,shuang2024conduction,bjerkelund1968high,boswell1983charge,mahy1982evidence,boswell1984charge,walker1985model,eaglesham1985microstructural,kusz1995modulated,rouxel2012crystal,tadaki1990electrical,bennett1991observation,bennett1992effects,butz2013nuclear,boswell2012advances,nataj2024raman,thompson2024solution,yu2024intrinsically} have furthermore attracted attention due to the simultaneous presence of various physical phenomena such as superconductivity \cite{yang2018pressure,yuan2020pressure}, magnetism \cite{petkov2023charge}, and multiple charge-density-wave (CDW) states. These various CDW states have been experimentally confirmed in electron diffraction \cite{eaglesham1985microstructural,boswell1983charge,mahy1982evidence}, Raman spectroscopy \cite{nataj2024raman}, X-ray diffraction (XRD) \cite{boswell1983charge,eaglesham1985microstructural,boswell1986structural,petkov2023charge,kusz1993incommensurate}, magnetisation \cite{petkov2023charge}, resistivity \cite{yang2018pressure,tadaki1990electrical,galvis2023nanoscale}, and scanning tunnelling microscopy (STM) \cite{galvis2023nanoscale,prodan1998scanning} measurements. These results were also thoroughly investigated theoretically \cite{walker1988nbte,morelli1989novel,morelli1989analysis,morelli1986transverse,chen1989superspace,goff1991order,whangbo1984band} and computationally \cite{galvis2023nanoscale,petkov2023charge}.

\begin{figure}[ht]
\includegraphics[width=\columnwidth]{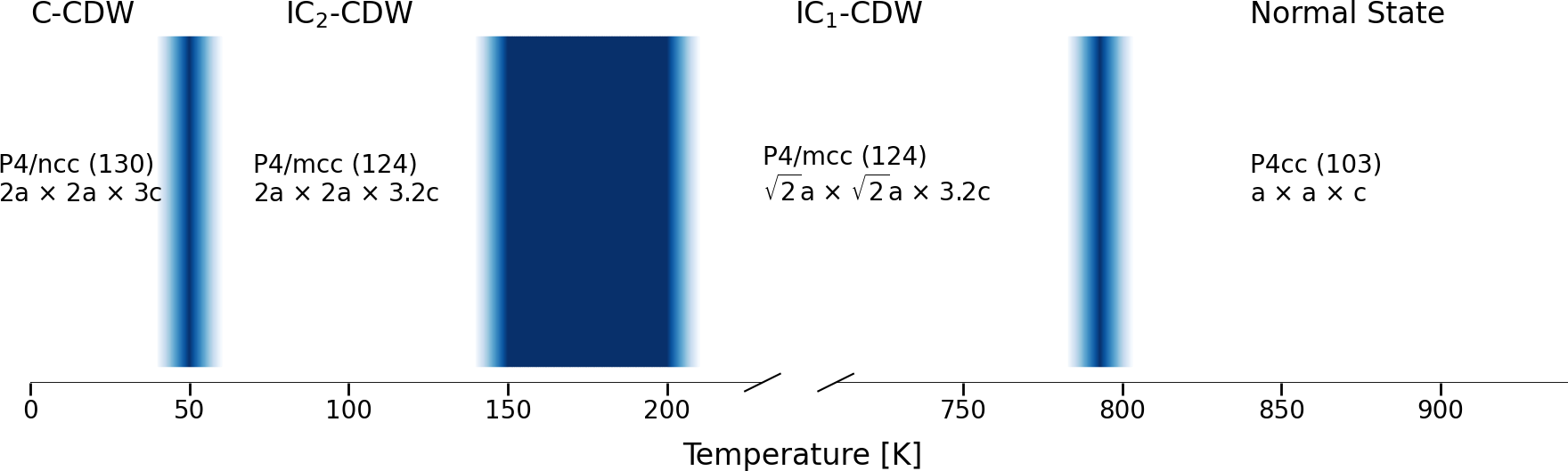}
\caption{Overview of the charge-density-wave states in NbTe$_4$. The crystal symmetry and CDW modulation changes as the temperature is varied. The temperature for the IC$_1$-CDW to IC$_2$-CDW transition varies among different publications between \SI{150}{\kelvin} and \SI{200}{\kelvin} \cite{tadaki1990electrical,walker1988nbte, petkov2023charge,yang2018pressure,kusz1993incommensurate,goff1991order}.}
\label{CDW_overview}
\end{figure}

An overview of the charge-density-wave transitions in NbTe$_4$ is shown in \autoref{CDW_overview}. Below around \SI{793}{\kelvin} the normal state obtains an incommensurate CDW (IC$_1$-CDW) modulation, which persists through room temperature \cite{bohm1987high} and acquires the crystal symmetry \textit{P4/mcc}. Lattice parameters in the unmodulated state were reported as \textit{a}$=\SI{6.499}{\angstrom}$, \textit{b}$=\SI{6.499}{\angstrom}$, and \textit{c}$=\SI{6.837}{\angstrom}$ \cite{galvis2023nanoscale} at room temperature. In this structure, Nb atoms are located at the centre of two square antiprisms of Te atoms, each rotated relative to the other. These components form linear chains along the \textit{c}-direction, connected to other chains by van der Waals interactions. As a consequence, single crystals of this compound grow as long needles along the \textit{c}-axis, highlighting the quasi-one-dimensional nature of its crystal structure.

\begin{figure}[ht]
\includegraphics[width=\columnwidth]{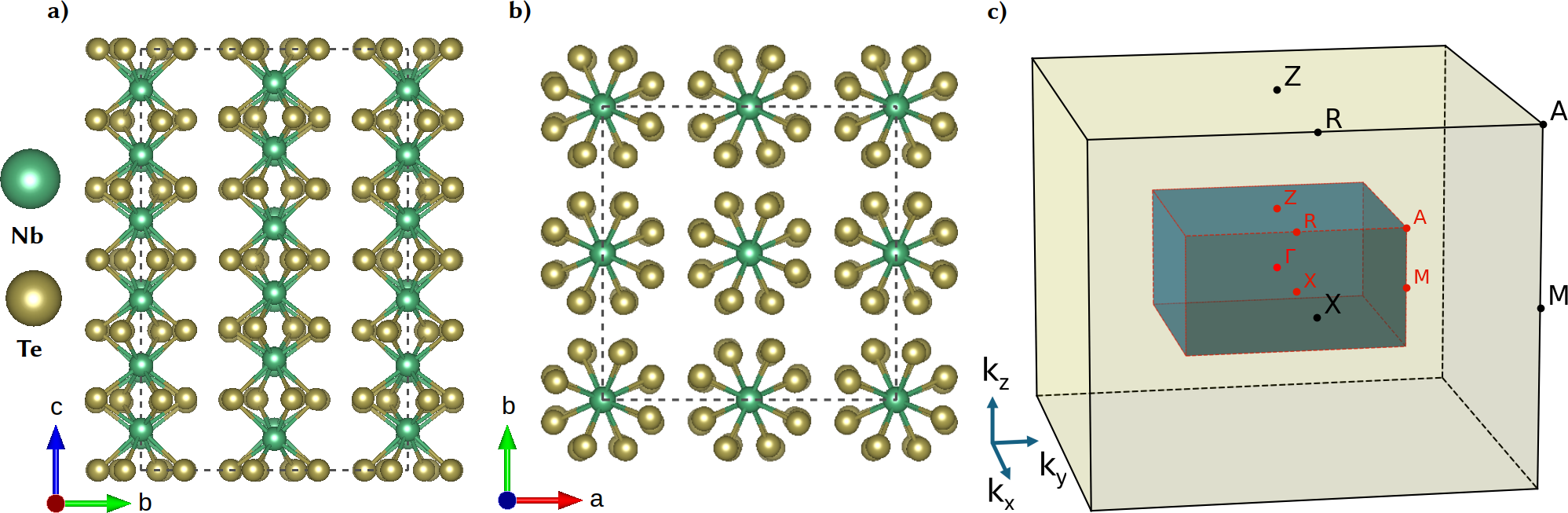}
\caption{a), b) Crystal structure and c) corresponding Brillouin zone of NbTe$_4$ in its commensurate charge-density-wave (C-CDW) state at temperatures below \SI{50}{\kelvin}. The crystal obtains a $2a\times2a\times 3 c$ supercell structure with space group \textit{P4/ncc}. The Brillouin zone of the unmodulated crystal is included for comparison.}
\label{crystal_structure}
\end{figure}

\begin{figure*}[t]
  \setkeys{Gin}{height=48mm}
  \subfloat{\includegraphics{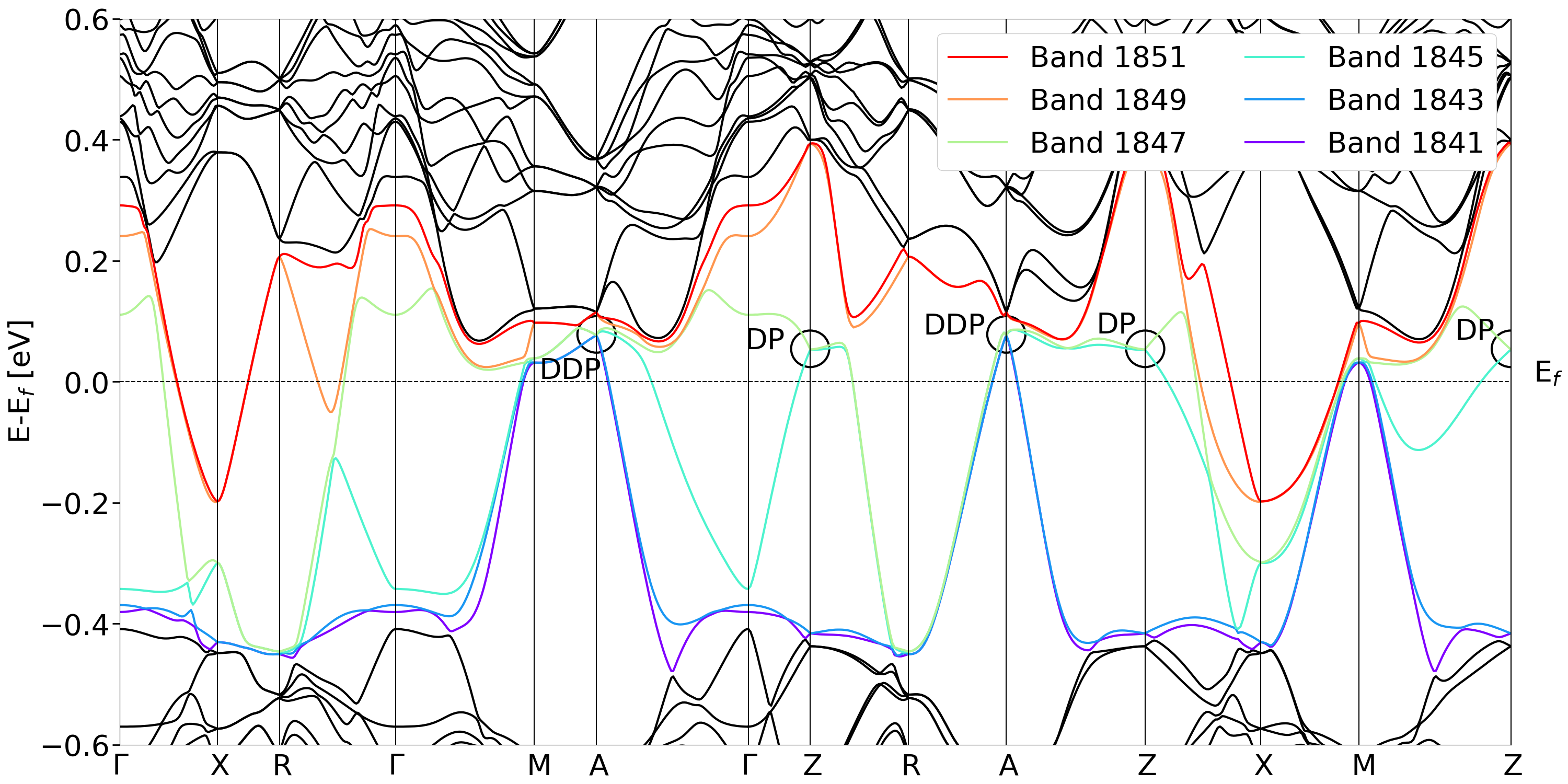}}
  \hfill
  \subfloat{\includegraphics{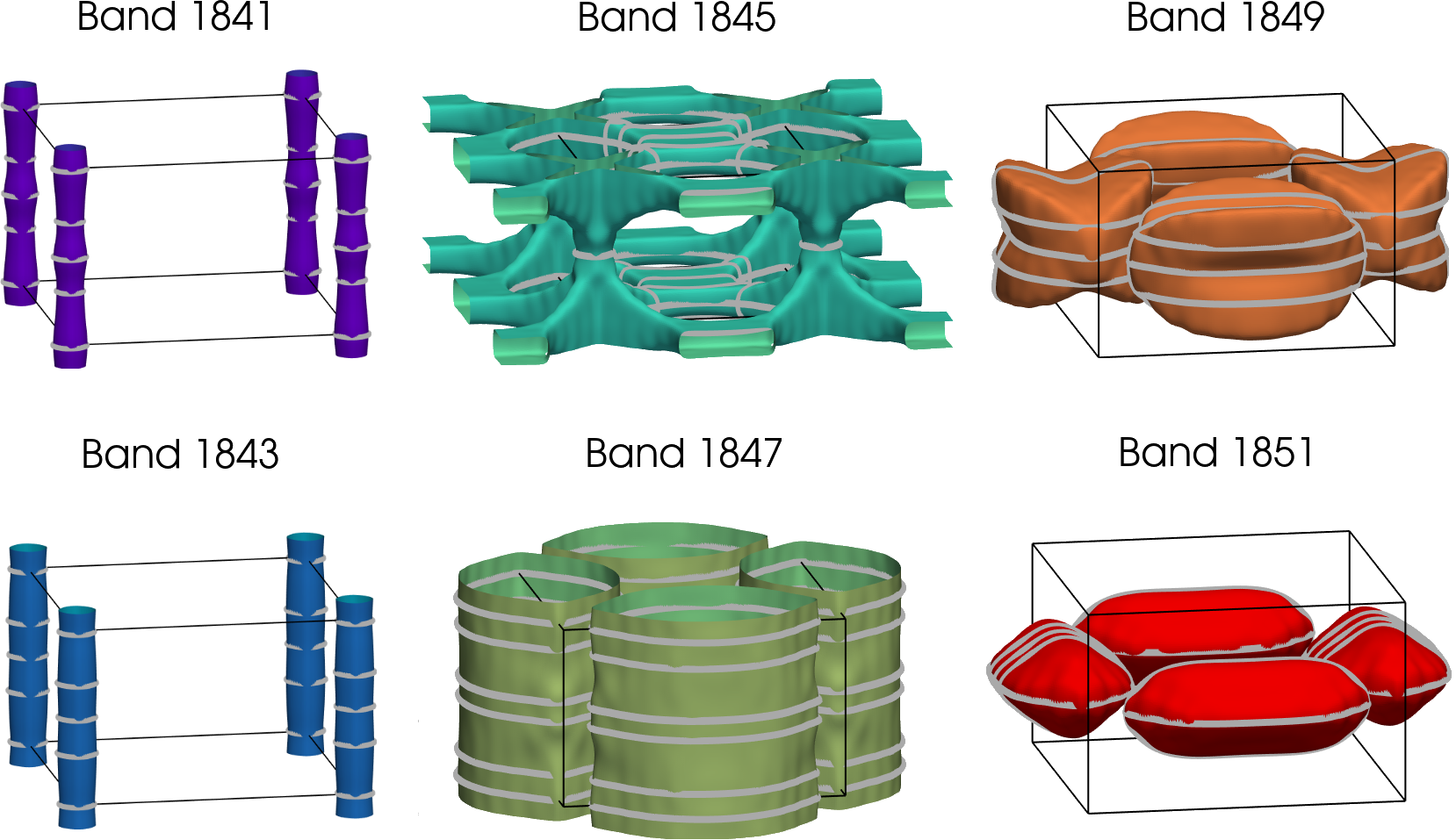}}
  \caption{Band structure in the commensurate charge-density-wave (C-CDW) state of NbTe$_4$, and corresponding Fermi surface extracted from those bands. Grey lines indicate extremal orbits around the individual Fermi sheets.}
  \label{band_structure}
\end{figure*}

Upon cooling down the crystal further, it undergoes a rare incommensurate-to-incommensurate charge-density-wave transition to the IC$_2$-CDW-state in the temperature range of \SI{150}{\kelvin}-\SI{200}{\kelvin}, changing the modulation while keeping its crystal symmetry. Finally, at around \SI{50}{\kelvin}, the crystal obtains a commensurate CDW (C-CDW) structure in the symmetry group \textit{P4/ncc} with a modulation of $2a\times2a\times 3 c$. The crystal structure and its corresponding Brillouin zone (BZ) are shown in \autoref{crystal_structure}, and the BZ of the unmodulated crystal structure is included for comparison.

Given this modulation and symmetry group at low temperatures, the band structure and its corresponding Fermi surface can be probed with experimental techniques such as ARPES or quantum oscillation measurements, both of which are absent in NbTe$_4$ in the literature. The work presented here provides insights into the electronic structure of this material by probing the Fermi surface with de Haas - van Alphen oscillations in the magnetic torque. The rotation study of such quantum oscillations confirms the presence of the commensurate charge-density-wave state at low temperatures and furthermore reveals magnetic breakdown orbits which can only be explained by tunnelling between electron and hole pockets in this material. These results are furthermore supported with transport measurements and DFT calculations, the latter of which indicates the occurrence of a rare eightfold degenerate double Dirac node.

\section{Methods}
\subsection{Numerical methods}
Density functional theory calculations shown here were obtained using Wien2k \cite{blaha2001wien2k}. The generalised gradient approximation (GGA) based on the PBE exchange-correlation potential \cite{perdew1996generalized} was used. The atomic sphere radii (muffin-tin radii) were chosen as $R_{mt} = 2.5$ a.u. for the Nb and Te atoms. The plane-wave cut-off parameter was chosen as  $R_{mt}K_{max} = 8$, where $R_{mt}$ is the smallest atomic sphere radius in the unit cell and $K_{max}$ is the magnitude of the largest $k$-vector. Calculations were performed on a $17 \times 17 \times 10$ k-point mesh in the full Brillouin zone. Convergence was reached for the charge ($<$ 0.0001 e), energy ($<$ 0.0001 Ry) and the interatomic force ($<$ 1.0 mRy/a.u.). Frequencies were extracted using the Supercell K-space Extremal Area Finder (SKEAF) \cite{rourke2009electronic,julian2012numerical}.
\subsection{Experimental methods}
Single crystals of NbTe$_4$ were grown via the self-flux growth method using Te as the flux. Nb powder (Aldrich Chemical Company: 99.9\%) and Te lumps (Alfa Aesar: 99.999\%) were mixed together with a molar ratio of Nb:Te = 1:8 inside a glove box filled with purified Argon gas (H$_2$O $<$ 0.5 ppm and O $<$ 0.5 ppm), and sealed in an evacuated quartz ampoule. The ampoule was then heated to \SI{1000}{\celsius} within \SI{24}{\hour} and held at that temperature for \SI{144}{\hour} before slowly cooling down to \SI{600}{\celsius} over the course of another \SI{144}{\hour}. NbTe$_4$, grows in shiny long needle-shaped single crystals as shown in \autoref{SEM}, and the energy dispersive X-ray spectrum (EDS) confirms that the atomic ratio is close to 1:4.

All measurements shown here were performed at a temperature of \SI{1.6(1)}{\kelvin} and at magnetic fields up to \SI{15}{\tesla} inside a customised Oxford instruments cryostat. Magnetotransport measurements were performed with a standard four-point measurement using a lock-in amplifier, while de Haas - van Alphen (dHvA) oscillations were observed in the magnetic torque of NbTe$_4$ single crystals measured with custom-made piezoresistive cantilevers.
\section{Theory}
\subsection{Band Structure}
The band structure in the C-CDW state is illustrated in \autoref{band_structure}. Bandfolding leads to an increase of the number of bands in the Brillouin zone, and in total six bands (highlighted in colour) and their respective spin-degenerate partners (labelled with even numbers, but not shown explicitly) cross the Fermi energy. The band structure shows a regular Dirac point (DP) at Z and an eightfold degenerate double Dirac point (DDP), a point at which four spin-degenerate bands cross, at A \cite{sambongi1993shubnikov,luo2017resistivity,gao2017anisotropic,zhang2020eightfold,rong2024dominant}. The existence of such double Dirac points is due to the commensurate charge density wave: the transformation at low temperatures turns the crystal structure into one of 7 among the 230 space groups, namely \textit{P4/ncc} (130), that allow for such band crossings \cite{wieder2016double}. NbTe$_4$ does not host such DDP at higher temperatures where it obtains the space group symmetries \textit{P4/mcc} (124) and \textit{P4cc} (103).

\subsection{Quantum Oscillations}
The Fermi sheets arising from the six bands crossing the Fermi energy are also included in \autoref{band_structure} with their respective extremal areas highlighted by grey lines. These can in principle be probed with quantum oscillation measurements in the magnetic torque $\tau$ via the relation \cite{shoenberg2009magnetic}

\begin{equation}\label{torque_LK}
	\resizebox{0.9\linewidth}{!}{$\tau = C B^{3/2}\sum\limits_{\text{i}} \frac{dF_i}{d\theta} \left|\frac{\partial^2 A_i}{\partial k^2_{\parallel}}\right|^{-1/2} \sum\limits_{p=1}^{\infty}p^{-3/2}R_T R_D R_S \sin\left(2\pi p \left(\frac{F_i}{B}-\gamma\right)\pm\delta\right)$}
\end{equation}
where $B$ is the magnetic field, $C$ is a $B$-independent constant, $\theta$ is the angle between magnetic field and the extremal area $A$, while $k_{\parallel}$ is the wavevector parallel to the magnetic field. The indices $i$ and $p$ range over all possible orbits and higher harmonics, respectively. $R_T$, $R_D$, and $R_S$ are various damping terms arising from finite temperature, finite electron lifetime and spin-splitting, respectively. The phase shift $\gamma$ arises from the topological properties of the semimetal, while $\delta$ depends on the dimensionality of the Fermi sheet. The frequency $F$ can be related to the extremal area $A$ via the Onsager relation:

\begin{equation}\label{Onsager_relation}
	F = \left(\frac{\hbar}{2\pi e}\right) A
\end{equation}
where $\hbar$ and $e$ are the reduced Planck constant and the electron charge, respectively. This treatment only takes fundamental orbits into account that are confined to the semiclassical motion on a particular Fermi sheet arising from the band structure in the material. However, for bands that are separated by only a small gap $\Delta k$ in reciprocal space, such motion can include multiple orbits simultaneously, which is known as magnetic breakdown (MB). The underlying mechanism here allows electrons/holes in a strong magnetic field to tunnel from one Fermi sheet to another, thus leading to an extremal area that is much larger than the one of its individual parts.

The dependence on the k-space gap in magnetic breakdown can be determined \cite{chambers1966magnetic} by considering that magnetic breakdown is exponentially suppressed with a probability given by
\begin{equation}\label{MB_criterion}
	P = \exp\left(-B_0/B\right) \quad \text{with} \quad B_0 = \frac{\pi}{2}\frac{\hbar}{e}\left(\frac{\Delta k^3}{a+b}\right)^{\frac{1}{2}}
\end{equation}
where $\Delta k$ describes the gap size measured at the closest point between two Fermi sheets in reciprocal space, while $a^{-1}$ and $b^{-1}$ are properties of the orbits and refer to the radii of curvature on each side of the gap. Magnetic breakdown is said to occur when this probability surpasses $1/e$ \cite{carter2010mean}, i.e. when the external field reaches the critical field, $B = B_0$. Note that usually multiple gaps need to be bypassed in order for the charge carriers to make a full revolution around the breakdown orbit. The corresponding critical magnetic field is therefore the sum of all the critical fields for each gap, i.e.
\begin{equation}
	B_c = n B_0
\end{equation}
where $n$ counts the number of gaps.

\section{Results and Discussion}

\begin{figure}[t!]
  \centering
  \includegraphics[width=\columnwidth]{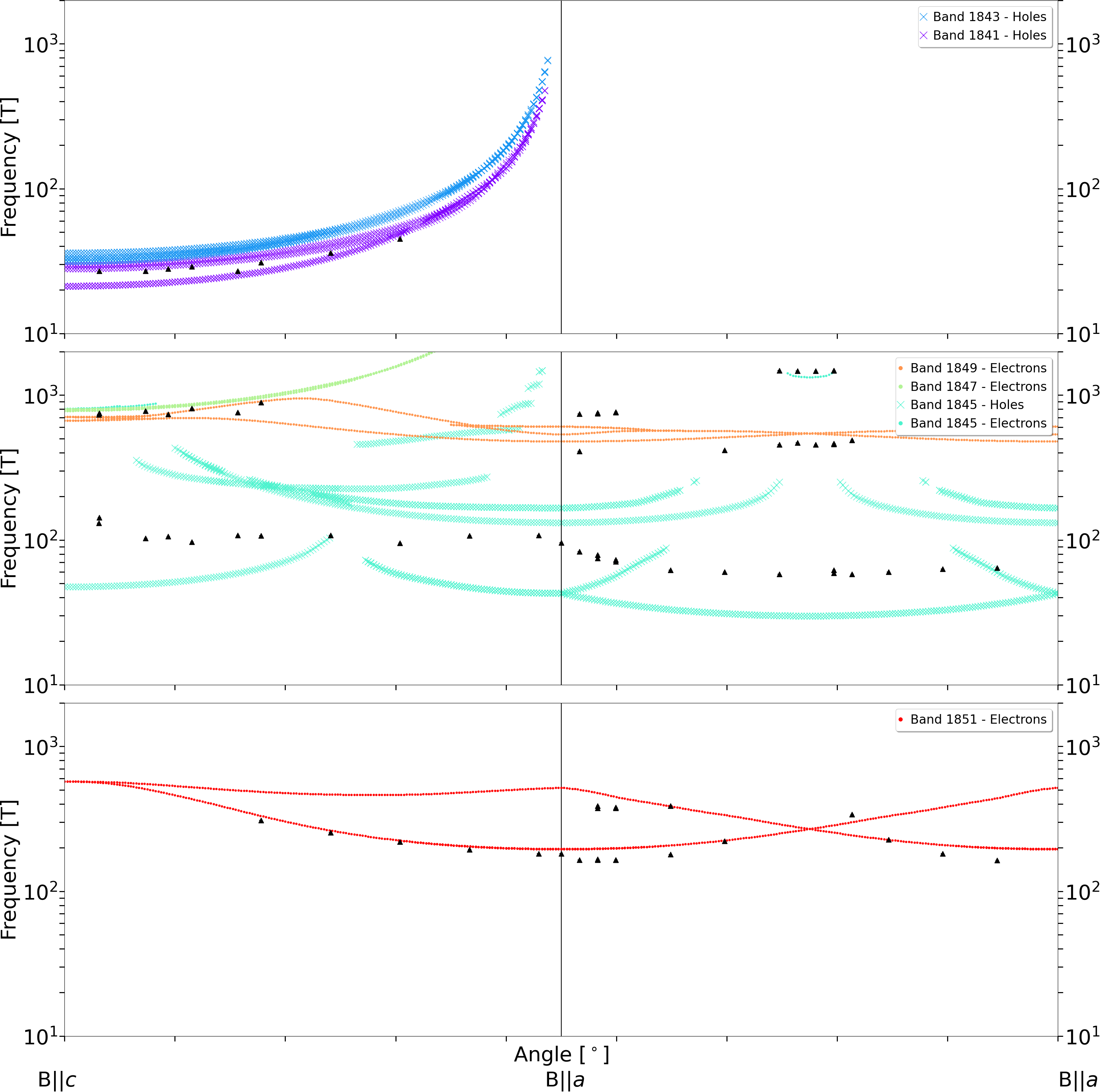}
  \caption{Extracted frequencies including measurement data (black triangles) in the \textit{c-a} and \textit{a-a} plane. Ticks on the x-axis correspond to steps of \SI{20}{\degree}. The DFT calculated frequencies arise from the commensurate \textit{P4/ncc} structure, whose Fermi surface is shown in \autoref{band_structure}. Raw Fast Fourier Transform data is shown in \autoref{FFT_NbTe4_1} and \autoref{FFT_NbTe4_2}}
  \label{NbTe4-C-CDW_2-P4ncc_frequencies}
\end{figure}

\begin{figure}[ht]
  \setkeys{Gin}{height=39mm}
  \subfloat{\includegraphics{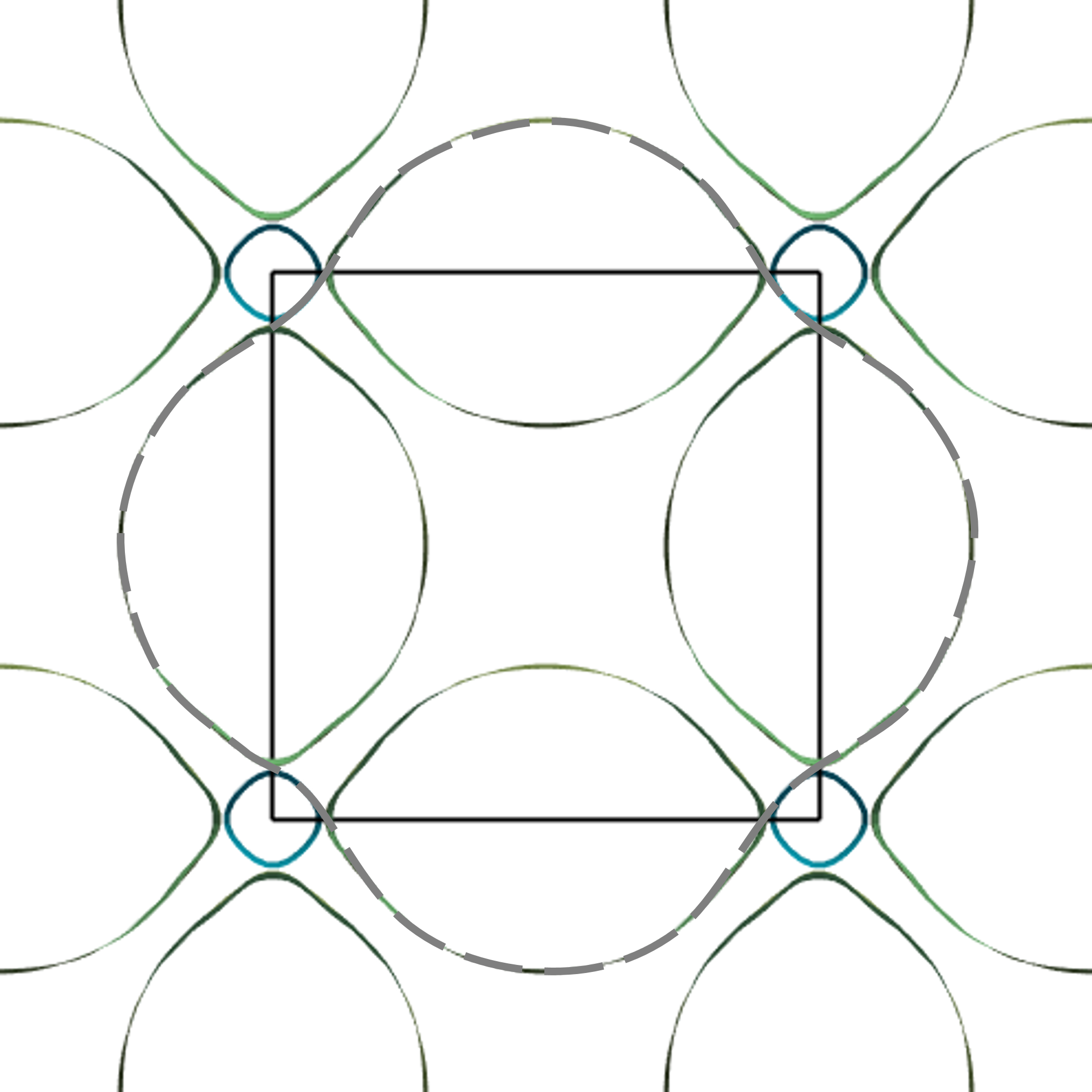}}
  \hfill
  \subfloat{\includegraphics{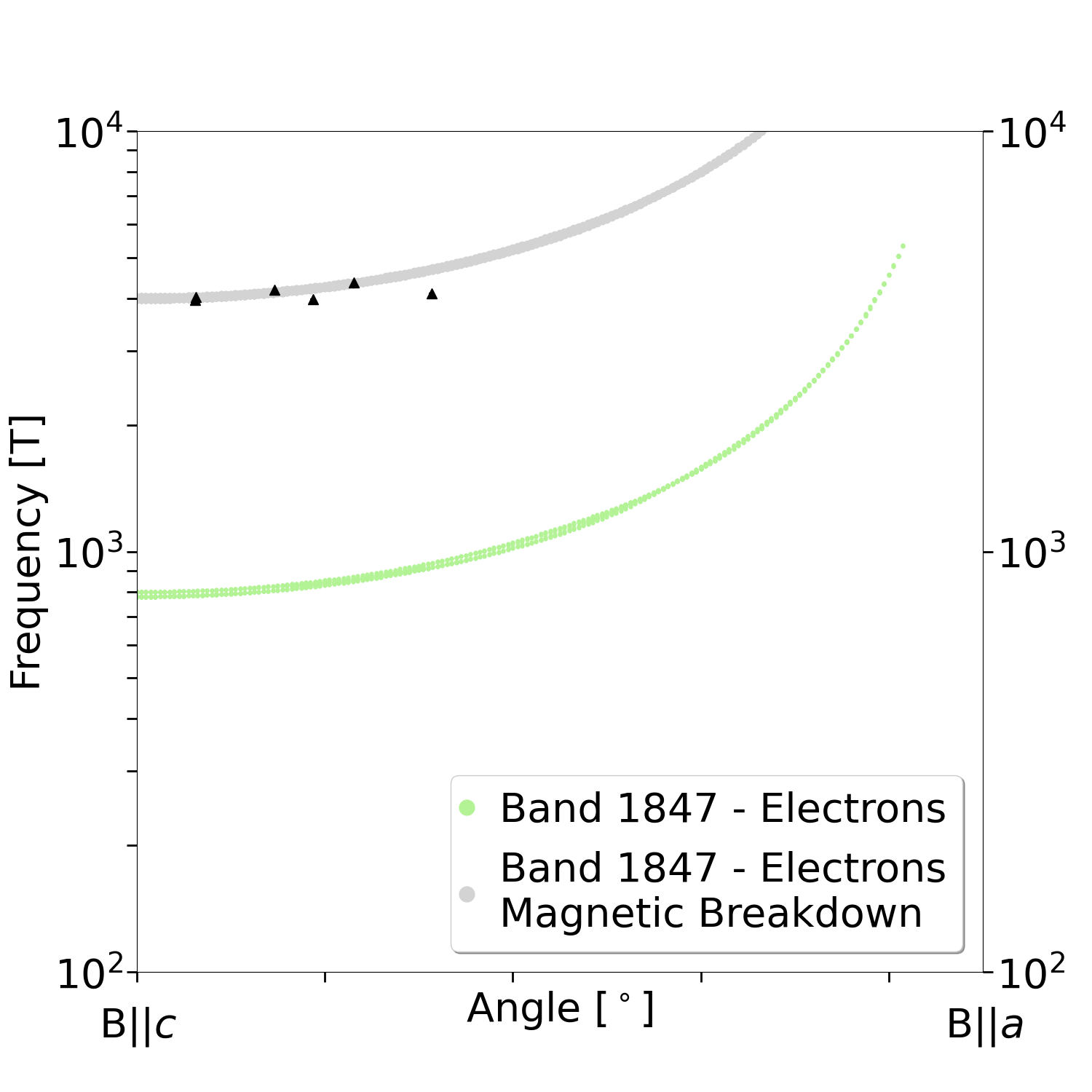}}
     \caption{Magnetic breakdown in NbTe$_4$. Left: Fermi surfaces of bands 1845 and 1847 close to the $k_z = 0$ plane as seen along the crystallographic \textit{c}-axis. Magnetic breakdown can introduce new frequencies through orbits that go around four lemon-shaped Fermi sheets (marked by a grey dashed line). Right: Measurement data (black triangles), and frequency branch of band 1847 (light green) in the \textit{c-a}-plane including the modified branch (grey) that arises from magnetic breakdown as described in the main text.}
        \label{NbTe4_magnetic_breakdown_orbits}
\end{figure}

The frequencies extracted from the extremal areas shown in \autoref{band_structure} are plotted in \autoref{NbTe4-C-CDW_2-P4ncc_frequencies} together with the dHvA magnetic torque data. For better visibility, the data points were distributed among the branches arising from the various Fermi sheets manually. A high frequency branch at around \SI{4}{\kilo\tesla} close to the B$\parallel$\textit{c}-axis was not included and will be discussed separately in \autoref{NbTe4_magnetic_breakdown_orbits}.

The thin cylinder-shaped Fermi surfaces arising from bands 1841 and 1843 can both explain data in the low frequency spectrum very well, however only one branch can be detected in the data, possibly due to the low purity of the crystal. The residual resistance ratio (RRR) in NbTe$_4$ was measured as RRR$\approx 6.5$, which is consistent with values in the literature \cite{tadaki1990electrical,yang2018pressure} and in stark contrast to its sister compound TaTe$_4$, which was reported with values of up to RRR=200 \cite{sambongi1993shubnikov}. The low purity in NbTe$_4$ might be one reason as to why quantum oscillation data has been absent in this material thus far.

The occurrence of a high-frequency branch around \SI{4}{\kilo\tesla} for fields aligned near the \textit{c}-axis cannot be explained with the DFT calculations of NbTe$_4$ in its commensurate CDW state with space group \textit{P4/ncc}. However, magnetic breakdown of the Fermi sheets present in this crystal can potentially explain the data. In particular band 1847 whose quasi two-dimensional lemon-shaped Fermi surface is shown in \autoref{NbTe4_magnetic_breakdown_orbits} appears to be a good candidate for magnetic breakdown, if it is assumed that electrons can tunnel to the Fermi sheets of band 1845 in between. This is justified since the orbits from band 1845 have hole-character, while the orbits from band 1847 have electron-character. The breakdown orbit will thus semiclassically appear as a clockwise rotation around the electron sheet, and as a counter-clockwise rotation around the hole-sheet \cite{van2018electron}. Furthermore, de Haas - van Alphen oscillations can be observed in both of these two Fermi sheets individually (see \autoref{NbTe4-C-CDW_2-P4ncc_frequencies}), indicating their clear presence in the magnetic torque.

To compute the overall frequency arising from such an orbit, we note that the branch originating from band 1845 starts at a frequency of around \SI{47}{\tesla}, while the branch originating from band 1847 starts at a frequency of around \SI{780}{\tesla} when B$\parallel$\textit{c}. For an orbit that encloses four lemon-shaped Fermi sheets at the edge of the Brillouin zone, we can then estimate the breakdown frequency by adding four halves of the lemon shaped Fermi sheet to the cross-sectional area of the Brillouin zone, while subtracting the cross-sectional area of four quarters of an orbit arising from band 1845. The area in reciprocal space can be calculated from the reciprocal lattice vectors, and the frequency can be related to that area using the Onsager relation in \autoref{Onsager_relation}. The overall frequency of the breakdown branch is then computed as \SI{3979}{\tesla} for B$\parallel$\textit{c}. As the magnetic field is tilted away from the \textit{c}-axis, the cross-sectional area of this enlarged magnetic breakdown orbit can be approximated as quasi two-dimensional in this angle-range and hence increases as \SI{3979}{\tesla}$/\cos{\theta}$. This is included in \autoref{NbTe4_magnetic_breakdown_orbits} and is consistent with the observed frequencies in the magnetic torque data which shows an upturn of the frequency branch as the magnetic field is tilted away from the \textit{c}-axis.

\begin{figure}[ht]
  \setkeys{Gin}{height=39mm}
  \subfloat{\includegraphics{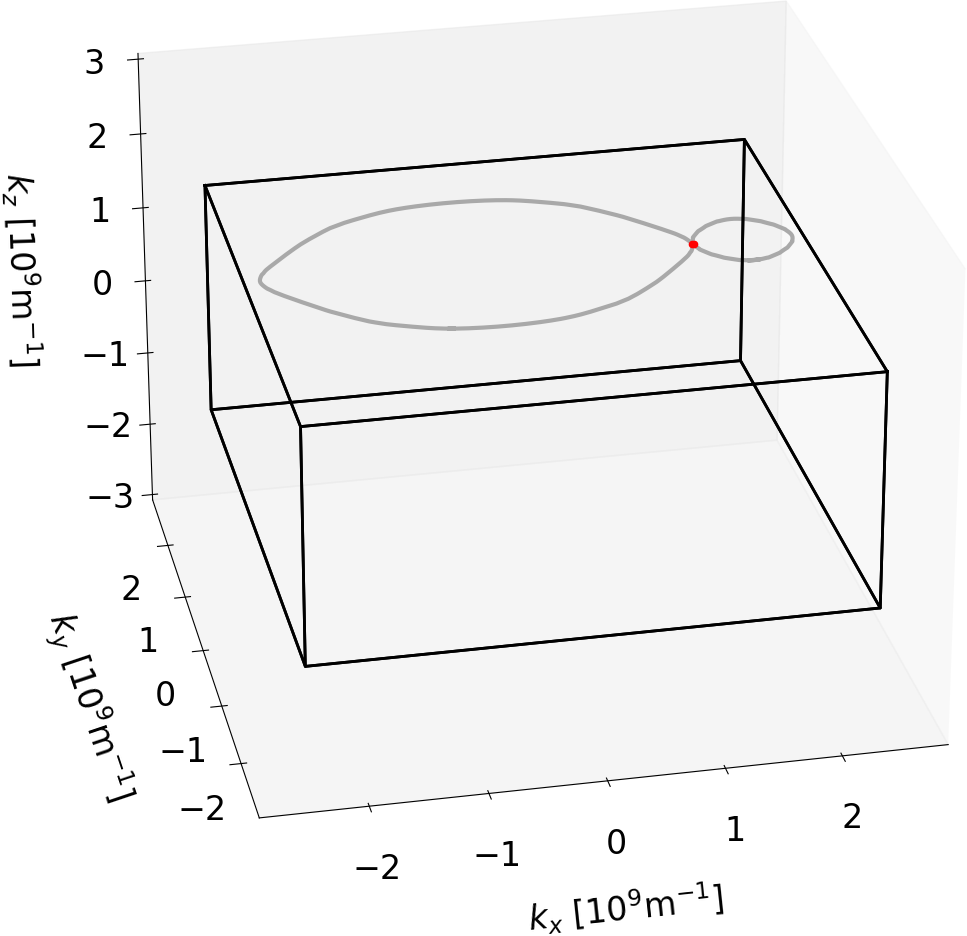}}
  \hfill
  \subfloat{\includegraphics{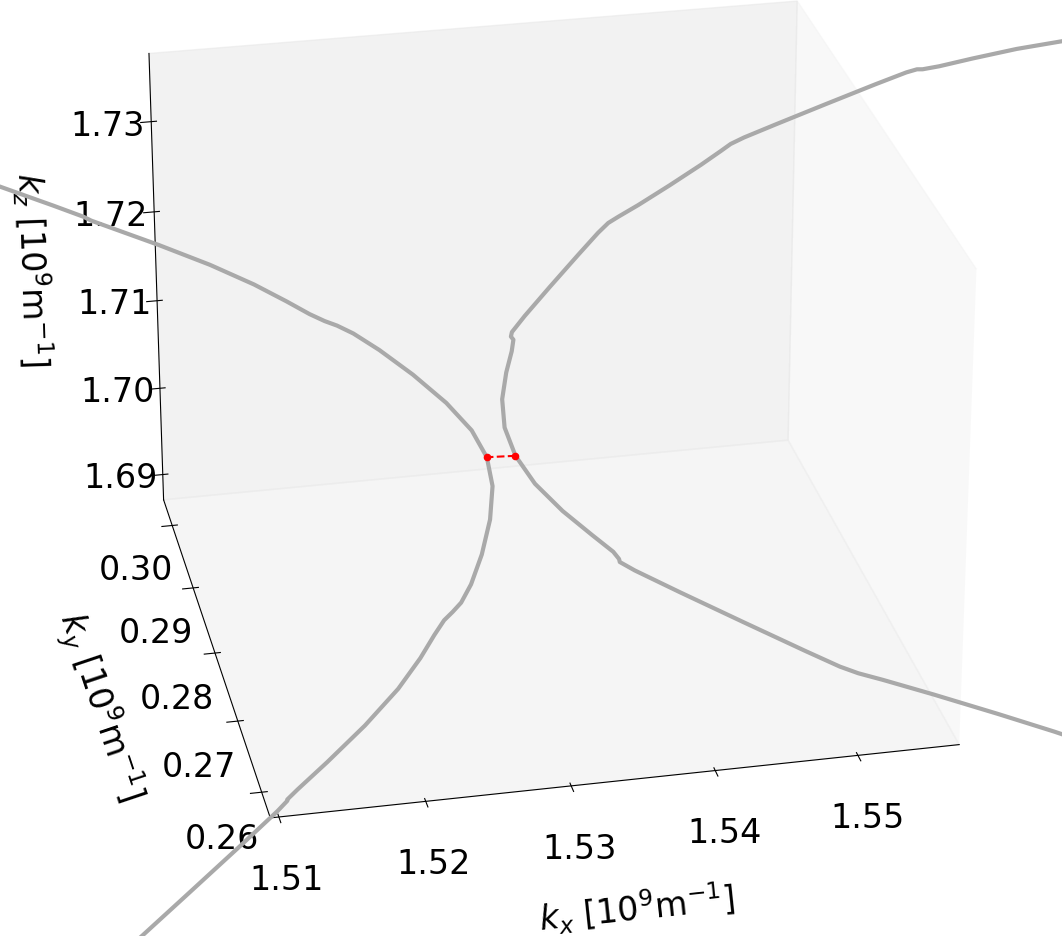}}
     \caption[Two orbits from bands 1845 and 1847 in NbTe$_4$ close to the $k_z = 0$ plane.]{Two orbits from bands 1845 and 1847 in NbTe$_4$ close to the $k_z = 0$ plane. The shortest distance between these two orbits is relevant in the condition for magnetic breakdown.}
        \label{NbTe4_MB_orbits}
\end{figure}

To see whether the condition in \autoref{MB_criterion} is fulfilled, the gap between two orbits from bands 1845 and 1847 are investigated in more detail in \autoref{NbTe4_MB_orbits}. The radii of curvature at the closest point are given by \SI{0.0250}{\per\angstrom} and \SI{0.0193}{\per\angstrom} for band 1845 and band 1847, respectively. The gap between the two orbits is \SI{0.0014}{\per\angstrom}, and hence the critical magnetic field needed for a charge carrier to cross this gap is given by $B_0 = \SI{0.6}{\tesla}$. For a charge carrier to make a full revolution around the breakdown orbit, eight gaps need to be crossed, resulting in a critical field of $B_c = \SI{4.8}{\tesla}$, which is well within the magnetic fields reached here.

\section{Conclusion}
We investigated the (double) Dirac semimetal NbTe$_4$, by probing de Haas - van Alphen oscillations as a function of the angle between applied magnetic field and the crystallographic axes of the material. Such quantum oscillation studies reveal an interband transition orbit at moderate magnetic fields of up to \SI{15}{\tesla}, which can be explained well within the framework of magnetic breakdown. Overall, the experimental study performed here agrees well with DFT calculated band structures, thereby indirectly confirming the presence of a commensurate CDW state with a $2a\times2a\times 3 c$ supercell structure in the space group \textit{P4/ncc}, and its corresponding Dirac nodal points. Given the insights provided here, future studies could focus on investigating the topological properties of this material in more detail.

\begin{acknowledgments}
We want to thank Darius-Alexandru Deaconu and Mohammad Saeed Bahramy for insightful discussions.

\end{acknowledgments}

\appendix

\section{Characterisation}\label{Characterisation}
\begin{figure}[ht]
  \setkeys{Gin}{height=26mm}
  \subfloat{\includegraphics{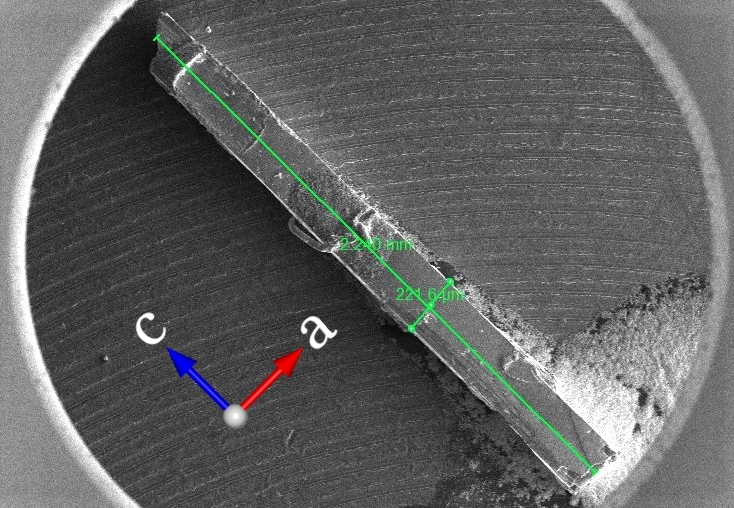}}
  \hfill
  \subfloat{\includegraphics{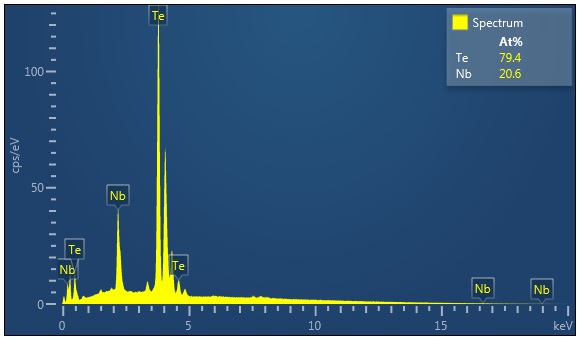}}
  \caption{Single crystals of NbTe$_4$ measured in a scanning electron microscope. Energy dispersive X-ray spectroscopy (EDS) shows an atomic ratio of nearly 1:4 as expected.}
  \label{SEM}
\end{figure}

\section{Transport properties}

\begin{figure}[ht]
  \includegraphics[width=0.49\columnwidth]{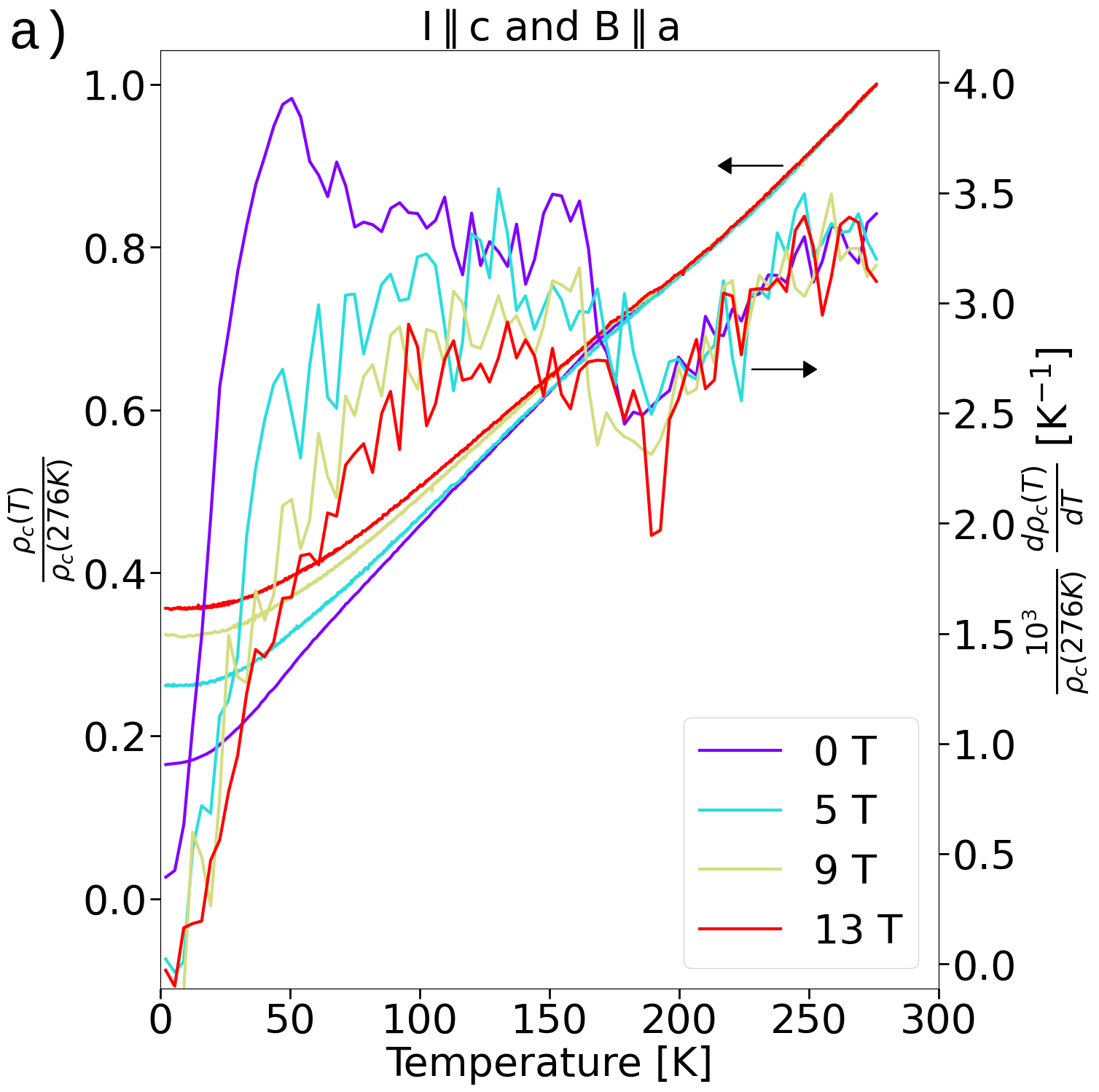}
  \includegraphics[width=0.49\columnwidth]{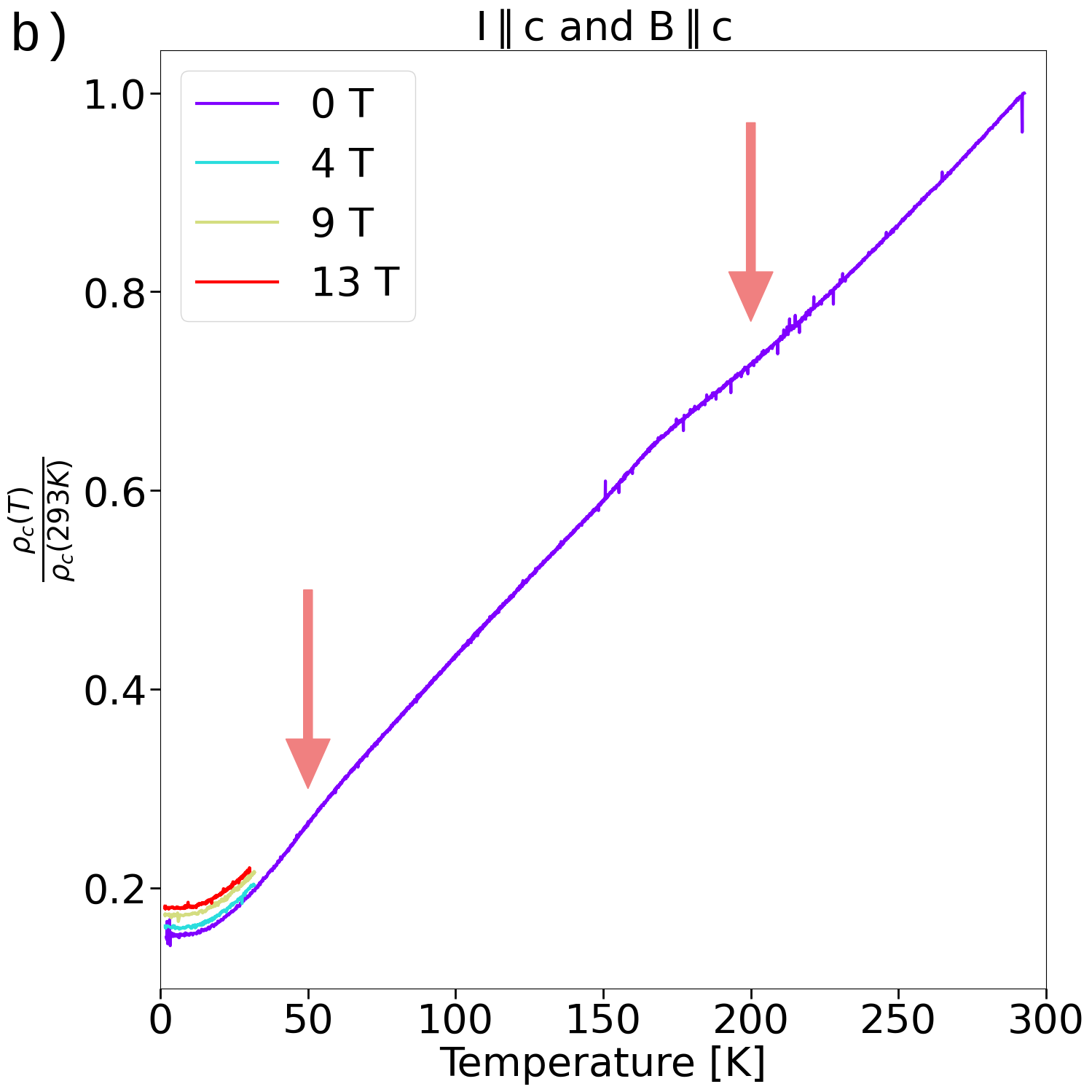}
  \includegraphics[width=0.49\columnwidth]{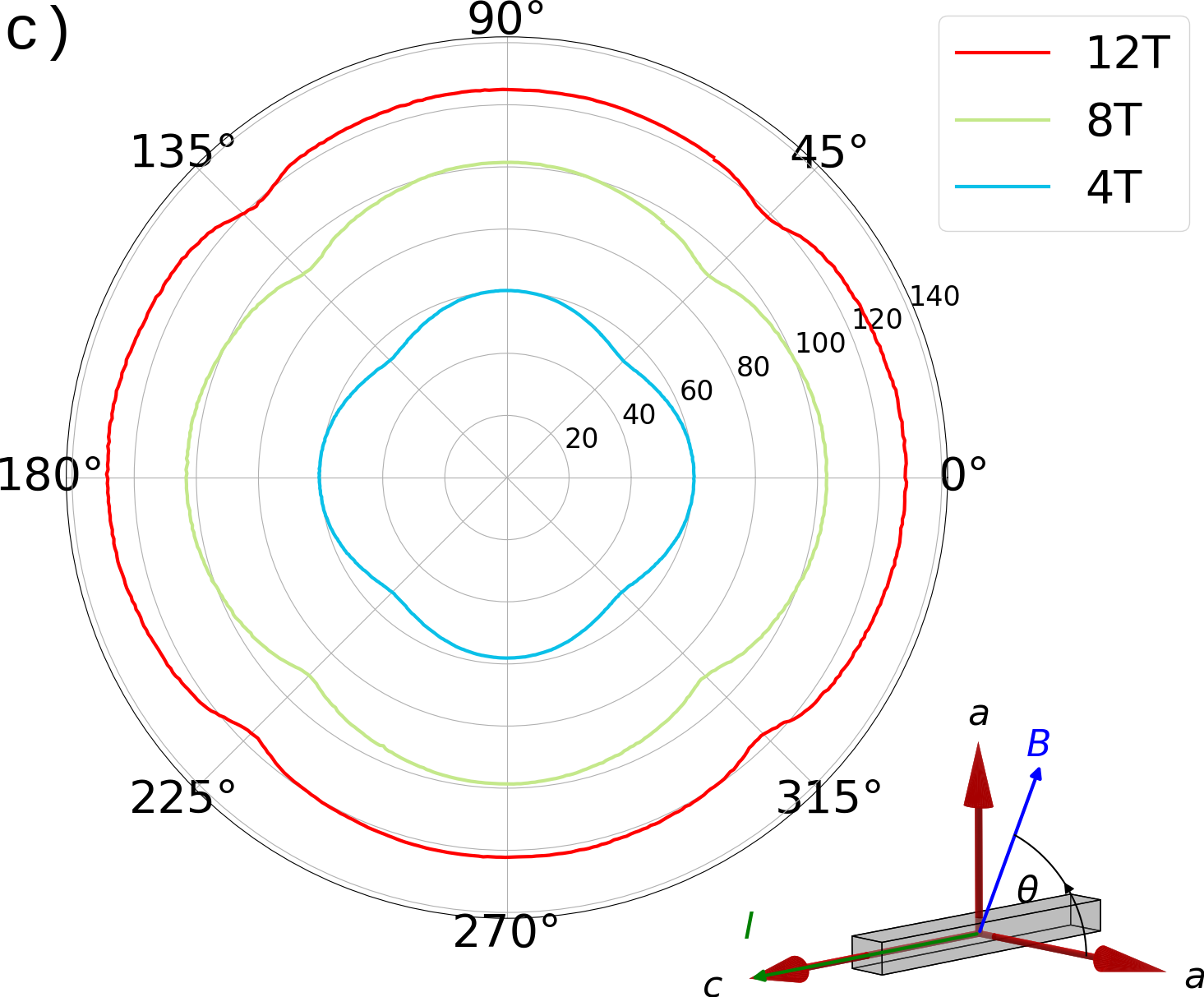}
  \includegraphics[width=0.49\columnwidth]{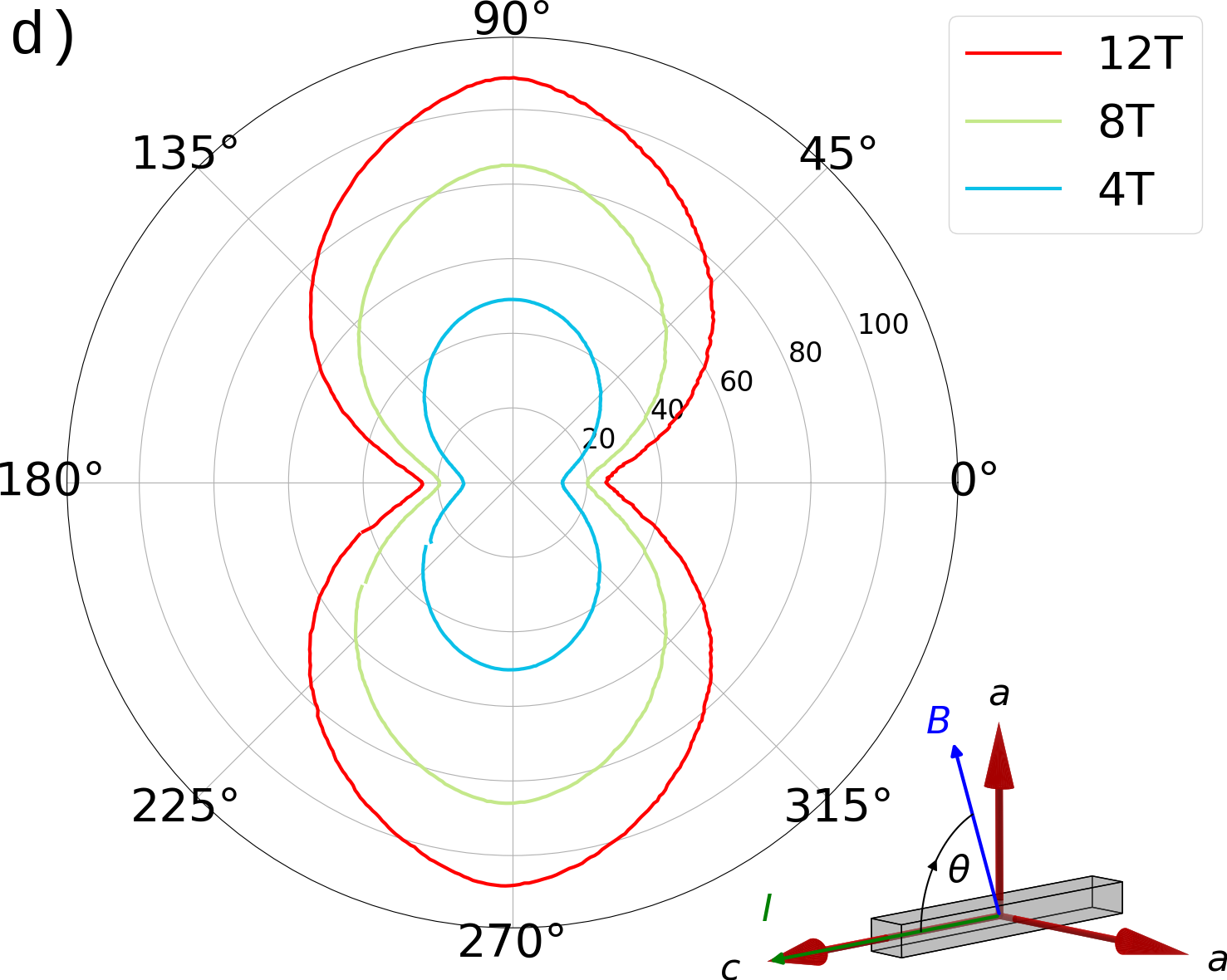}
  \caption{a), b): Cooldown curves in field for a NbTe$_4$ single crystal with the current applied along the crystallographic \textit{a}-axis and the magnetic field parallel to either the \textit{a}-axis or \textit{c}-axis. Red arrows indicate the IC$_1$-CDW to C-CDW transition and the IC$_1$-CDW to IC$_2$-CDW transition. c), d): Angular magnetoresistance (AMR) of a NbTe$_4$ single crystal rotated in a static magnetic field.}
  \label{transport}
\end{figure}

\begin{figure}[ht]
  \centering
  \includegraphics[width=\columnwidth]{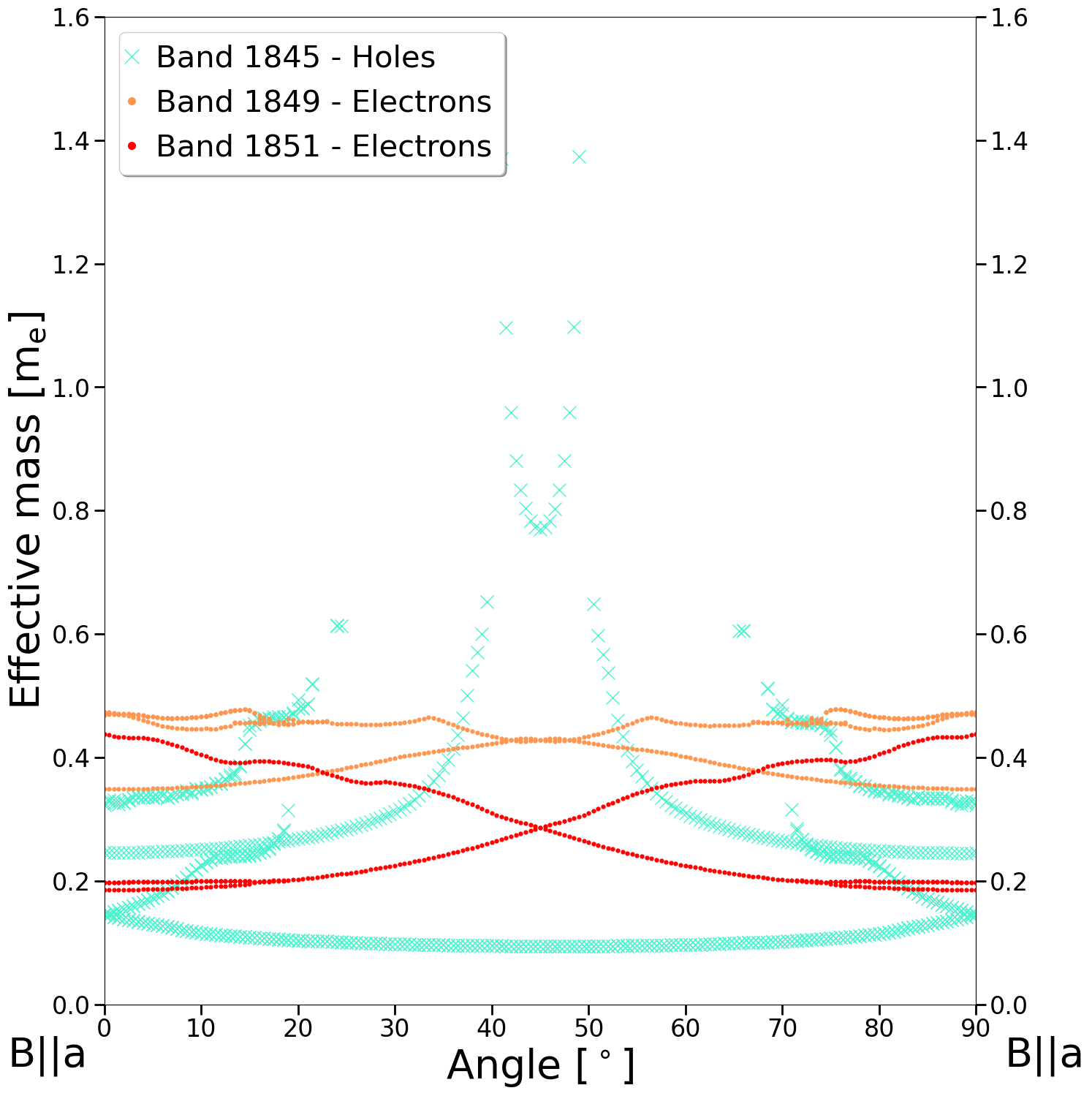}
  \caption{Effective masses as a function of angle for NbTe$_4$ in the \textit{a-a} plane extracted from DFT calculations for the commensurate CDW state with the symmetry group \textit{P4/ncc}.}
  \label{appendix_NbTe4_eff_masses_a-b}
\end{figure}

The temperature-dependent resistivity at various static fields is illustrated in \autoref{transport} a) and b). The derivative for the data shown for B$\parallel$\textit{a} was determined by interpolating the raw data with a cubic spline. Two kinks are observed, indicating the IC$_1$-CDW to C-CDW transition and the IC$_1$-CDW to IC$_2$-CDW transition. Magnetic fields applied along this crystallographic axis suppress the IC$_1$-CDW to C-CDW transition. Furthermore, a resistivity upturn can be seen, but any attempt to fit the data with Kohler's rule was unsuccessful, similar to what was reported by Yang et al. \cite{yang2018pressure}. For measurements with B$\parallel$\textit{c}, no significant deviation from the zero field resistivity can be observed.

Angular magnetoresistance (AMR) measurements shown in \autoref{transport} c) and d) illustrate the four-fold (\textit{a-a} plane) and two-fold (\textit{a-c} plane) periodicity of the crystal structure, respectively. Overall, with the current applied along the crystallographic \textit{c}-axis, the magnetoresistance defined in terms of the resistivity $\rho(B)$ as
\begin{equation}\label{semiclassical_MR}
	\text{MR} = \frac{\rho(B)-\rho(0)}{\rho(0)}
\end{equation}
reaches its highest values of $\sim 125\%$ when the magnetic field is aligned with the crystallographic \textit{a}-axis. The angular magnetoresistance in the \textit{c-a} plane shows strong anisotropy with two-fold symmetry in a peanut-like shape. The AMR with the magnetic field in the \textit{a}-\textit{a} plane shows a dip in the data at around \SI{45}{\degree} which has also been observed in TaTe$_4$ \cite{gao2017anisotropic}, albeit much more pronounced in the latter. This dip in the resistivity is reminiscent of the discussion of the Dirac nodal-line semimetal ZrSiS \cite{ali2016butterfly} where it could be connected to a topological phase transition. Although a topological origin in NbTe$_4$ is possible, the weak angular dependence more resembles the AMR of NbIrTe$_4$ \cite{schonemann2019bulk} for which it was argued that the angular magnetoresistance is a result of the anisotropy of the Fermi surface and hence of the effective masses. Indeed, such anisotropy can be observed in the corresponding DFT calculations, for the effective masses as shown in \autoref{appendix_NbTe4_eff_masses_a-b}. The effective masses in the \textit{a-a} plane for band 1845 approximately double as they approach \SI{45}{\degree}, which can in principle lead to the dip in the AMR observed here.

\section{Raw FFT Data}

\begin{figure}[ht]
  \centering
  \includegraphics[width=\columnwidth]{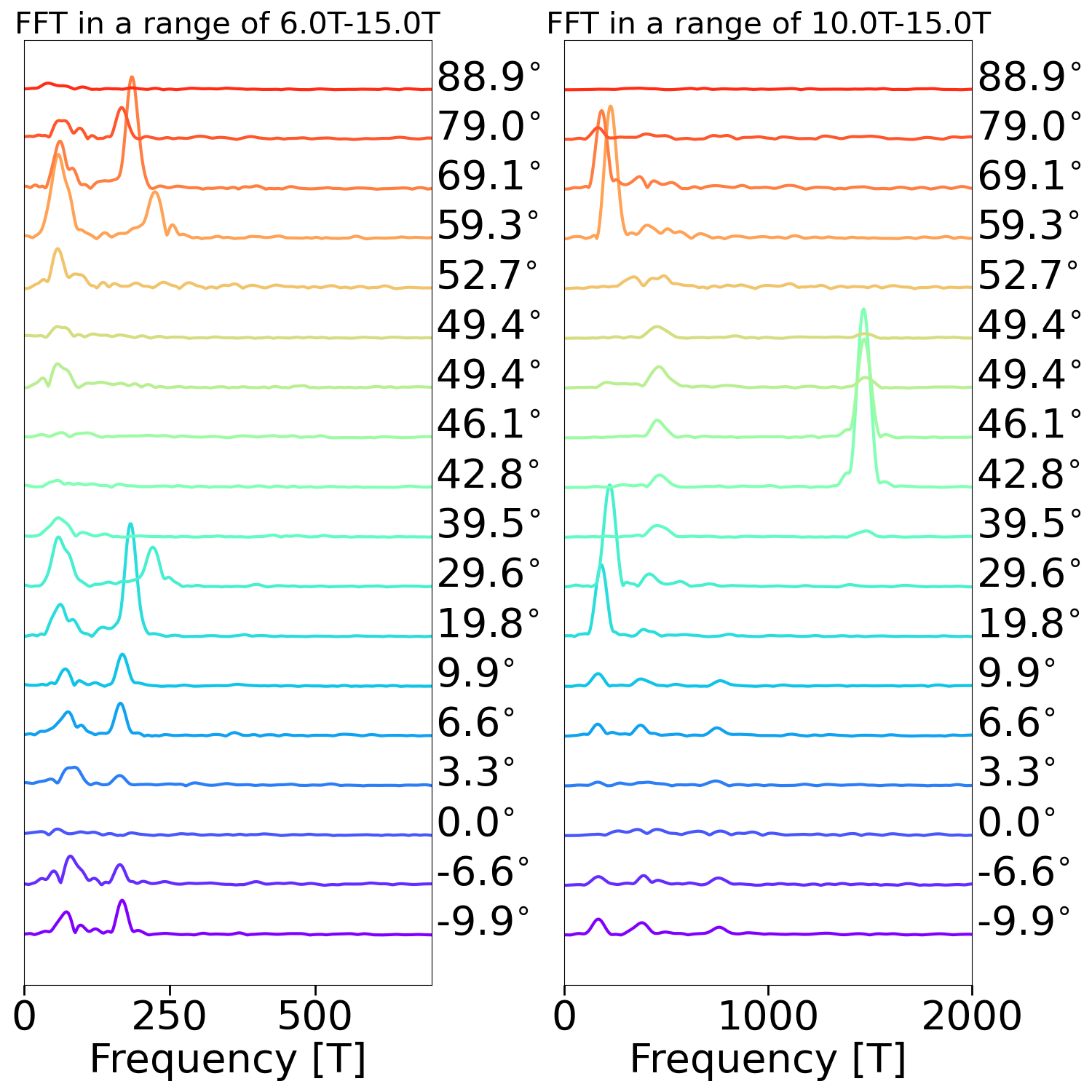}
  \caption{Raw Fast Fourier Transform for the data shown in \autoref{NbTe4-C-CDW_2-P4ncc_frequencies} for NbTe$_4$ in the \textit{a-a} plane. \SI{0}{\degree} corresponds to the magnetic field being aligned with the \textit{a}-axis.}
  \label{FFT_NbTe4_1}
\end{figure}

\begin{figure}[ht]
  \setkeys{Gin}{height=70mm}
  \subfloat{\includegraphics{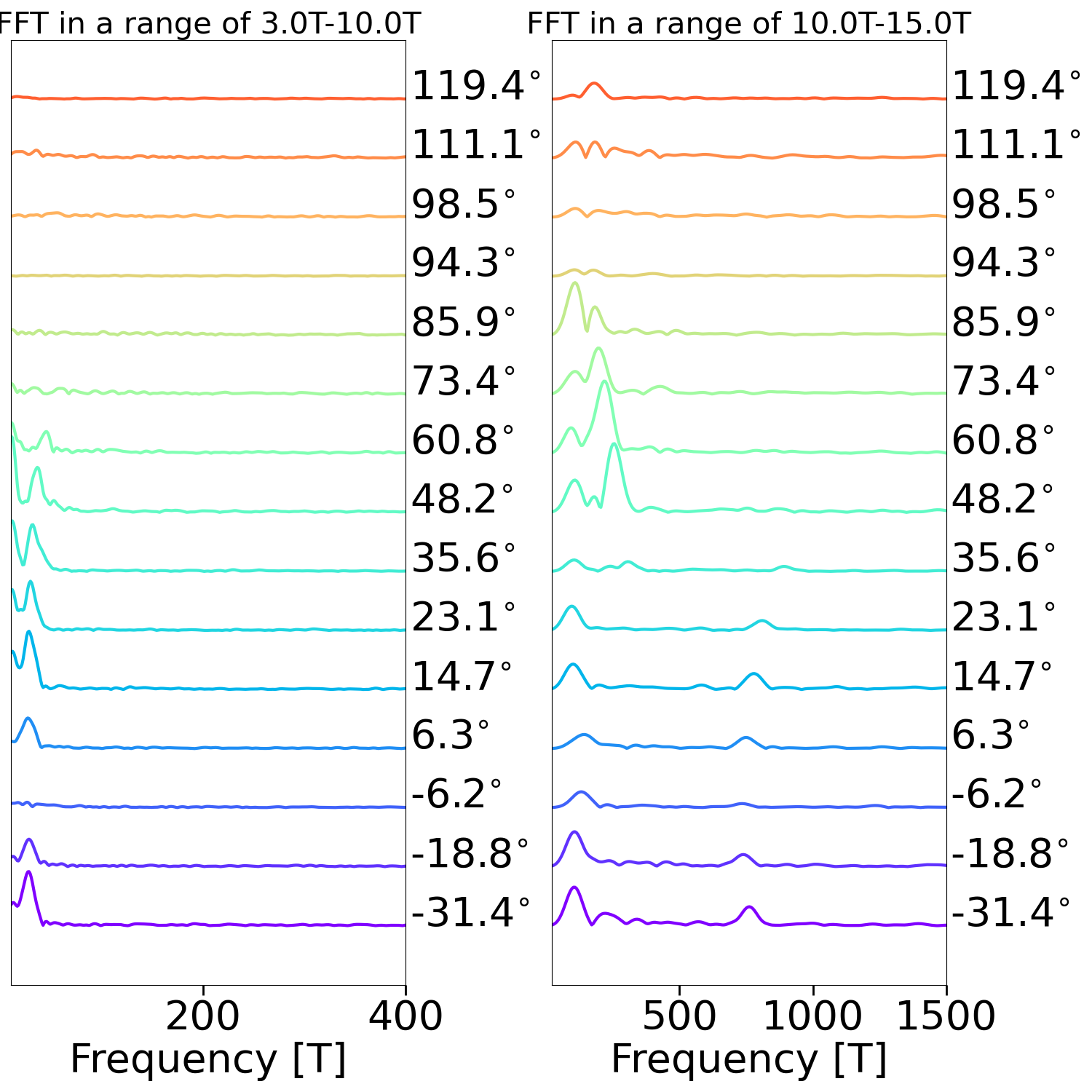}}
  \hfill
  \subfloat{\includegraphics{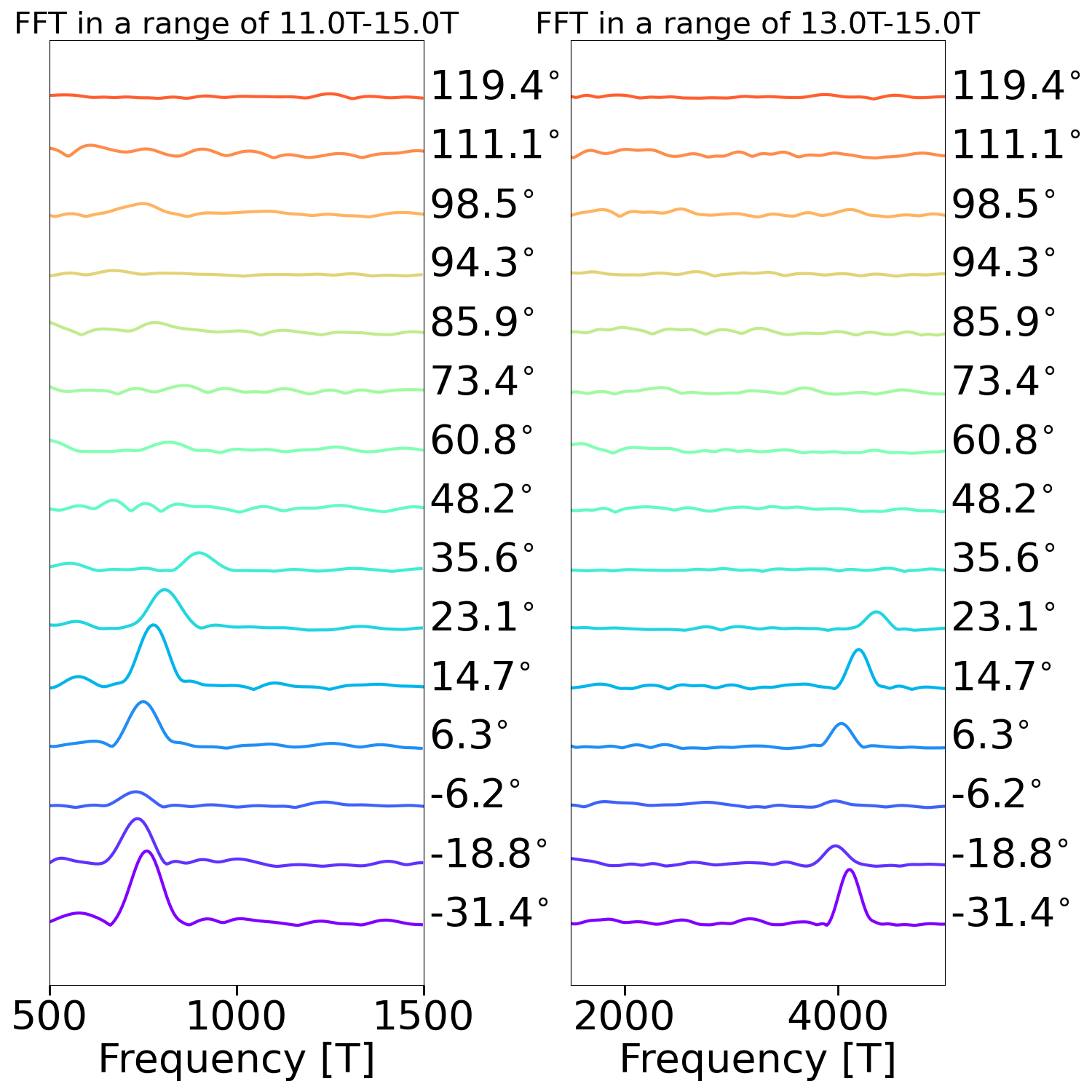}}
     \caption{Raw Fast Fourier Transform for the data shown in \autoref{NbTe4-C-CDW_2-P4ncc_frequencies} for NbTe$_4$ in the \textit{c-a} plane. \SI{0}{\degree} corresponds to the magnetic field being aligned with the \textit{c}-axis.}
     \label{FFT_NbTe4_2}
\end{figure}

\newpage
\pagebreak
\clearpage

\onecolumngrid

\bibliographystyle{unsrt}  
\bibliography{NbTe4_QO.bib}

\end{document}